\documentclass[conference]{IEEEtran}
\IEEEoverridecommandlockouts
\usepackage{amsfonts}
\usepackage{amsmath}
\usepackage{balance}
\usepackage{cite}
\usepackage{acronym}

\usepackage{hyperref}

\usepackage{epstopdf}
\usepackage{epsf}

\usepackage{ifpdf}
\ifCLASSINFOpdf
   \usepackage[pdftex]{graphicx}
   \DeclareGraphicsExtensions{{.pdf}}
\else
   \usepackage[dvips]{graphicx}
   \DeclareGraphicsExtensions{{.eps}}
\fi

\usepackage{pifont}
\newcommand{\cmark}{\ding{51}}%
\newcommand{\xmark}{\ding{55}}%

\usepackage[usenames]{color}

\acrodef{BRN}[BRN]{Barrage relay network}
\acrodef{CBR}[CBR]{controlled barrage region}
\acrodef{OLA}[OLA]{opportunistic large array}
\acrodef{MANET}[MANET]{mobile ad hoc network}
\acrodef{TDMA}[TDMA]{time division multiple access}
\acrodef{GPS}[GPS]{Global Positioning System}
\acrodef{RTS}[RTS]{request-to-send}
\acrodef{CTS}[CTS]{clear-to-send}
\acrodef{SNR}[SNR]{signal-to-noise ratio}
\acrodef{SINR}[SINR]{signal-to-interference-plus-noise ratio}
\acrodef{CCI}[CCI]{co-channel interference}
\acrodef{TC}[TC]{transport capacity}
\acrodef{MLSE}[MLSE]{maximal-likelihood sequence estimation}

\begin{document}
\abovedisplayskip=0.16pt
\belowdisplayskip=0.16pt
\hyphenation{multi-symbol}
\title{Unicast Barrage Relay Networks: \\ Outage Analysis and Optimization}
\author{Salvatore Talarico\IEEEauthorrefmark{1}, Matthew C. Valenti\IEEEauthorrefmark{1}, and Thomas R. Halford\IEEEauthorrefmark{2} \\
\IEEEauthorblockA{\IEEEauthorrefmark{1}West Virginia University, Morgantown, WV, USA.\\
\IEEEauthorrefmark{2} TrellisWare Technologies, Inc., San Diego, CA, USA.
 }
\vspace{-0.5cm}
}
\date{}
\maketitle

\thispagestyle{empty}

\begin{abstract}
\emph{Barrage relays networks} (BRNs) are ad hoc networks built on a rapid cooperative flooding primitive as opposed to the traditional point-to-point link abstraction. \emph{Controlled barrage regions} (CBRs) can be used to contain this flooding primitive for unicast and multicast, thereby enabling \emph{spatial reuse}. In this paper, the behavior of individual CBRs is described as a Markov process that models the potential cooperative relay transmissions.  The outage probability for a CBR is found in closed form for a given topology, and the probability takes into account fading and co-channel interference (CCI) between adjacent CBRs.  Having adopted this accurate analytical framework, this paper proceeds to optimize a BRN by finding the optimal size of each CBR, the number of relays contained within each CBR, the optimal relay locations when they are constrained to lie on a straight line, and the code rate that maximizes the transport capacity.
\end{abstract}

\section{Introduction}

A significant research investment has been made over the past decade on the topic of {\em cooperative communications} for wireless networks in general, and \acp{MANET}  in particular.  The improvements provided by it have been widely noted in the literature \cite{scaglione2006,stankovic2006}.  See, for instance, \cite{Sendonaris2003,Sendonaris2003b,Laneman2004,scaglione2003,chugg:2006} for example protocols involving single-antenna mobiles that cooperatively transmit and/or receive.
{\em \acp{BRN}} are a kind of cooperative \ac{MANET} designed for use at the tactical edge \cite{halford:2010a}. BRNs utilize time division multiple access (TDMA) and cooperative communications as the basis of an efficient flooding protocol wherein packets ripple out from sources in pipelined spatial waves. In a BRN, simultaneous transmissions of the same packet are not suppressed but rather exploited for the resulting diversity gains. The spatial extent of unicast and multicast transmissions can be contained in BRNs via {\em \acp{CBR}}. Briefly, a CBR is established by identifying a ring of buffer nodes around a portion of the network containing a source and its destination(s). Within the CBR, the barrage flooding primitive is used to transport data. The buffers suppress their relay function, thereby enabling multiple CBRs to be active at the same time. The increase in network capacity afforded by this \textit{spatial reuse} was analyzed in \cite{halford:2012} under standard information theoretic assumptions.

BRNs resemble the \acp{OLA} introduced by Scaglione and Hong in \cite{scaglione2003}. Indeed, analogs to CBRs in OLAs have been proposed \cite{ingram:2009}. Both OLAs and BRNs exploit the diversity that can be obtained when multiple nodes transmit identical packets. BRNs can be distinguished from early OLA descriptions by the time synchronization and receiver signal processing employed in BRNs which enables packets to be longer and data rates to be larger than the inverse of the maximum relative delay spread between cooperating transmitters. The most important difference between more recent OLA descriptions (e.g., \cite{ingram:2010}) and BRNs is the latter's use of \textit{autonomous} cooperative communications \cite{chugg:2006} as opposed to distributed space-time coding. This feature makes BRNs more suited for use in highly dynamic, tactical environments.

In this paper, unicast data transport in a CBR is modeled as a Markov process. Since relaying in BRNs is entirely opportunistic, a Markov chain is used to track every possible cooperative link transmission. The transition probabilities are evaluated for a given network topology via a closed-form expression for the outage probability of each cooperative transmission \cite{Talarico2013}, which takes into account path loss, Rayleigh fading, noise, and \ac{CCI} from adjacent CBRs.  A key analytical challenge is that on the one hand, CCI influences which nodes successfully receive and therefore affects the transition probabilities of the Markov process, yet on the other hand, the transition probabilities of the Markov process determine which nodes transmit and therefore affect the interference.  This coupling between the transition and outage probabilities is solved using an iterative approach, whereby the transition probabilities of the local \ac{CBR} are used to seed the transition probabilities of neighboring \acp{CBR}, assumed to be similarly configured. On the best of our knowledge, Markov processes were also used to analyze a cooperative system in other papers, but the interference was neglected, e.g \cite{marchenko2014,Hassan:2011}, or, when it is considered, simplified assumptions, which are accurate only for high density networks, were made \cite{Jung:2014} in order to facilitate the analysis.

Having established an analysis that can obtain exact expressions for the throughput of a given network configuration, this paper proceeds to optimize the network with respect to the {\em \ac{TC}}, which is a measure of forward progress.  In particular, the optimal \ac{CBR} size is identified, along with the optimal configuration (number and location) of relays within each \ac{CBR}, and the optimal code rate for each transmission.


\begin{figure*}[ht]
\centering
\includegraphics[width=14.25cm]{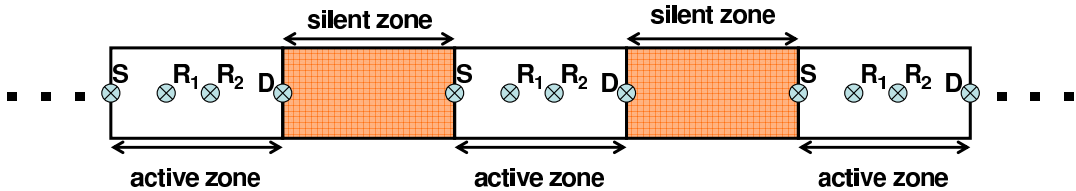}
\vspace{-0.30cm}
\caption{Example of \acp{BRN}, composed of multiple \acp{CBR}. Each \ac{CBR} is composed a four-node network.}
 \label{Example_BNR}
 \vspace{-0.50cm}
\end{figure*}

\section{Barrage Relay Network} \label{Section:BRN}

A \ac{BRN} is a simplified abstraction of a tactical MANET architecture that is currently being used in the field \cite{halford:2010a,halford:2012}. BRNs employ TDMA and cooperative communications.  All nodes utilize a common  frame format that requires coarse slot-level synchronization. This can be accomplished via a distributed network timing protocol when a satellite reference is unavailable. BRNs use autonomous cooperative communications rather than distributed beamforming or space-time coding to (i) minimize the communications overhead required for cooperation and (ii) enable multiple transmitters to participate in cooperative links to multiple destinations without knowing which other nodes are transmitting. As detailed in \cite{chugg:2006}, each node applies an independent, random phase-dithering pattern to each relayed message. This induces a time-varying fading characteristic at each receiver. By changing the phase dither \textit{within} a packet, destructive interference across an entire packet can be avoided. When modern codes with long block lengths are used, the phase-dithering technique effectively translates spatial diversity from multiple transmitters into time diversity that can be captured via \ac{MLSE} equalization and iterative detection.

The \ac{BRN} flooding primitive works as follows. During the first slot of a frame, the source broadcasts its message.  During the second slot, any node that received and successfully decoded the initial transmission will rebroadcast it. If more than one node concurrently transmits the message, then the superposition of their signals is received.
The diverse signal components are effectively maximal-ratio combined at each receiver. During each subsequent slot, all messages that were successfully decoded during the previous slot are rebroadcast. The process repeats until either the destination is reached, no receiver successfully decodes a transmission, or a maximum number of transmissions is reached.   Packets thus propagate outward from the source via a \emph{decode-and-forward} approach.  To prevent relay transmissions from propagating back towards the source, each node relays a given packet only once.

\section{Controlled Barrage Regions} \label{Section:CBR}

In  a  traditional  network,  a  link  is  defined  by  a  transmit/receive  radio  pair  that  share  a  suitably  reliable  point-to-point communications  channel. A  unicast route  is simply a  series  of  these  links  connecting  a  source-destination  node pair. In a \ac{BRN}, however, links are cooperative and comprise multiple transmitters and receivers. A cooperative path is made up of series of cooperative links between the source and destination nodes. A \ac{CBR} is simply the union of one or more such cooperative paths within some subregion or {\em zone} of the overall network.   \acp{CBR} afford  a  mechanism  for  unicast transport  that  is  more  robust  than  both  traditional  unicast routing and non-cooperative multipath variants \cite{halford:2011}.

\acp{CBR} can be established by specifying a set of \emph{buffer} nodes around a set of cooperating \emph{interior} nodes. Each buffer node acts as the source for transmissions into a \ac{CBR} and the destination for transmissions emanating from within the \ac{CBR}.  External packets may enter the \ac{CBR} only through a buffer node, and packets internal to a \ac{CBR} may only propagate to the rest of the network through a buffer node.  In  this  way,  multiple unicast transmissions may be established in different portions of the network. This is referred to as \emph{spatial reuse}. The buffer and interior nodes corresponding to a given source-destination  pair can be specified via the broadcast of \ac{RTS} / \ac{CTS} packets \cite{halford:2011}.

An example of linear \ac{BRN} is shown in Fig. \ref{Example_BNR}.  Each \ac{CBR} in the example is made up of four nodes. The two interior nodes of each \ac{CBR}, $R_1$ and $R_2$, act as relays, while the other two nodes are buffers, acting as the source $S$ and destination $D$ of the transmissions within the \ac{CBR}, respectively.  In the case of a linear BRN, each \ac{CBR} is defined to be a segment of length $d$, here assumed to be the same for all \acp{CBR}.

Let $N$ be the number of interior nodes in a \ac{CBR}. It is assumed that the transmission from a \ac{CBR}'s source to its destination must take place within one radio frame or else an {\em outage} occurs\footnote{Note that in practice, BRNs often employ a fixed frame length of $F=4$ and a fixed maximum number of relays that is greater than $F$ \cite{halford:2010a}. In such networks, spatial pipelining enables multiple packets to be active within a given CBR at the same time. In the interest of analytical tractability, this paper considers only interference from transmissions in different CBRs. This is motivated by the observations that (i) in practice, most tactical unicast traffic is local so that CBRs will typically be less that $4$ hops long and (ii) inter-CBR interference is more problematic than intra-CBR interference when $F=4$.}. Since each node transmits each packet at most once, the maximum number of transmissions per frame is $F=N+1$.  Since each transmission requires one slot, the maximum number of slots per radio frame is also $N+1$.  While variable-length frames could be supported, this would complicate the synchronization of neighboring \acp{CBR}. It is thus assumed that all \acp{CBR} have $N$ interior nodes  and $F=N+1$.



A packet may be broadcast by the source of a \ac{CBR} during only the first time slot of a frame.  In ideal channel conditions, it is possible that the destination receives the initial source transmission in the first slot, though more generally, additional transmissions from the interior nodes may be required for successful reception.  The buffer node acting as the source for one \ac{CBR} also acts as the destination for the previous \ac{CBR}.   Because the nodes operate in a half-duplex mode, it is not possible for the buffer to simultaneously transmit into one \ac{CBR} while receiving from another \ac{CBR}.  Thus, it is advisable to only allow one of the two \acp{CBR} serviced by a given buffer node to be active; i.e., either the \ac{CBR} that the buffer node is transmitting into is active or the \ac{CBR} that the buffer node is receiving from is active, but not both.  More generally, this suggests that \acp{CBR} can be divided into two types during a particular radio frame:
1) {\em active zones} and 2) {\em silent zones}.   During a given radio frame, only the nodes in active zones may transmit.  As illustrated in Fig. \ref{Example_BNR}, zones will typically alternate between active zones and silent zones, though it is feasible to have even further spatial separation between active zones.  A key advantage to alternating between active and silent zones is that it provides spatial separation between \ac{CCI}, as the \ac{CCI} in a given active zone is due to transmissions in other active zones, which are at least two zones away.
%

\section{Network Model} \label{Section:SystemModel}

Consider a \ac{BRN} composed of $K$ \acp{CBR}. The \ac{BRN} is a set of $M$ mobile radios or \emph{nodes} $\mathcal X = \{X_1, ..., X_M\}$.  The variable $X_i$ represents both the $i^{th}$ node and its location.  In the following, $X_i$ is typically used to denote a transmitting node and $X_j$ used to describe a receiving node.
During the $t^{th}$ time slot of a frame, $t\in\{1,...,F\}$, node $X_i$ transmits with probability $p_i^{(t)}$.  Any node in a silent zone will be characterized by $p_i^{(t)}=0$ for all slots of the frame.   Since the source of an active CBR may only transmit in the first slot of the frame, it is characterized by $p_i^{(1)}=1$ and $p_i^{(t)}=0$ for $t \in \{2, ..., F\}$.

Let $\mathcal X_j^{(t)} \subset \mathcal X$ be the set of cooperating or {\em barraging} nodes that transmit identical packets to $X_j$ during the $t^{th}$ time slot, $|\mathcal X_j^{(t)}|$ the number of barraging transmitters, and $\mathcal G_j^{(t)}$ the set of the indexes of the nodes in $\mathcal X_j^{(t)}$.
Nodes within the same \ac{CBR} as $X_j$ but not in $\mathcal X_j^{(t)}$ do not transmit; i.e., there is no intra-CBR interference.  However, nodes in other active \acp{CBR} are potential sources of interference, and each such node will generate \ac{CCI} with probability $p_i^{(t)}$.  The value of $p_i^{(t)}$ for each of these nodes will depend on the dynamics of the protocol, as will be described shortly.


When $X_i$ transmits, it broadcasts a signal whose average received power in the absence of fading is $P_{i}$ at a reference distance $d_{0}$.  For ease of exposition, it is assumed that all the $X_i$ that transmit do so with a common power $P_i = P$.  However, this assumption could be relaxed at the expense of slightly complicating the notation and analysis.  $X_i$'s power at receiver $X_j$ during time slot $t$ is
\begin{eqnarray}
  \rho_{i,j}^{(t)}
  & = &
  P g_{i,j}^{(t)} f( d_{i,j} ) \label{eqn:power}
\end{eqnarray}
where  $g_{i,j}^{(t)}$ is the power gain due to fading, $d_{i,j} = ||X_i - X_j||$ is the distance from $X_j$ to $X_i$, and $f( \cdot )$ is a path-loss function.
The $\{ g_{i,j}^{(t)} \}$ are independent and exponentially distributed with unit mean,
corresponding to Rayleigh fading.
It is assumed that the \{$g_{i,j}^{(t)}\}$ remain fixed for the duration of a time slot, but vary independently from slot to slot. For $d\geq d_{0}$, the path-loss function is
expressed as the attenuation power law
\begin{eqnarray}
   f \left( d \right)
   & = &
   \left( \frac{d}{d_0} \right)^{-\alpha} \label{eqn:pathloss}
\end{eqnarray}
where $\alpha > 2$ is the attenuation power-law exponent, and $d_0$ is sufficiently large that the signals are in the far field.

A commonly accepted approach to analyzing  diversity combining is  to  assume that the interference realizations in the different branches are the same; i.e., the interference is fully-correlated among the branches (cf., \cite{Aalo2000}).
While strictly speaking, this condition is not always met, the assumption makes the analysis manageable.  Furthermore, as a worst-case scenario, the analysis makes it possible to obtain an upper bound on outage probability, which happens to be tight \cite{Tanbourgi2013}.   Under this assumption and by using (\ref{eqn:power}) and (\ref{eqn:pathloss}), the instantaneous \ac{SINR} at mobile $X_j$ during slot $t$ is
\begin{eqnarray}
   \gamma_j^{(t)}
   & = &
   \frac{ \displaystyle \sum_{k \in \mathcal G_j^{(t)}} g_{k,j}^{(t)} \Omega_{k,j}  }{ \displaystyle \Gamma^{-1} + \sum_{i \not\in \mathcal G_j^{(t)}} I_i^{(t)} g_{i,j}^{(t)} \Omega_{i,j} }
   \label{Equation:SINR2}
\end{eqnarray}
where $\Gamma = d_0^\alpha P/\mathcal{N}$ is the \ac{SNR} of a unit-distance transmission when fading is absent, $\Omega_{i,j} = d_{i,j}^{-\alpha}$ is the relative path gain, $I_i^{(t)}$ is a Bernoulli variable indicating the $X_i$ is a source of interference during slot $t$, and $P[I_i^{(t)}=1]=p_i^{(t)}$.

Let $\beta$ denote the minimum \ac{SINR} required by $X_j$ for reliable reception and $\boldsymbol{\Omega }_j=\{\Omega_{1,j},...,\Omega _{M,j}\}$ represent the set of relative path gains from all $\{ X_i\}$ to $X_j$.  An \emph{outage} occurs when the \ac{SINR} falls below $\beta$. Conditioning on the path gains $\boldsymbol{\Omega }_j$ and the set of barraging nodes $\mathcal X_j^{(t)}$, the outage probability of mobile $X_j$ during slot $t$ is
\begin{eqnarray}
   \epsilon_j^{(t)}
   & = &
   P \left[ \gamma_j^{(t)} \leq \beta \Big| \boldsymbol \Omega_j, \mathcal X_j^{(t)} \right].
   \label{Equation:Outage1}
\end{eqnarray}
Because it is conditioned on $\boldsymbol{\Omega }_j$, the outage probability depends on the particular network realization, which has dynamics over timescales that are much slower than the fading. From \cite{Talarico2013}, the outage probability in Rayleigh fading  is
\begin{eqnarray}
 \epsilon_j^{(t)} \hspace{-0.25cm}&=& \hspace{-0.25cm}
1 - \hspace{-0.1cm}\sum_{k \in \mathcal G_j^{(t)}} \hspace{-0.1cm} \exp\left(-\frac{\beta}{\Omega_{k,j} \Gamma }\right) \prod_{s \in \mathcal G_j^{(t)},s\neq k} \frac{\Omega_{k,j}}{\Omega_{k,j}-\Omega_{s,j}} \nonumber \\
\hspace{-0.25cm} & & \hspace{-0.25cm}\times \prod_{ i \notin \mathcal G_j^{(t)} } \frac{\Omega_{k,j}+\beta\left( 1-p_i^{(t)} \right) \Omega_{i,j}}{ \Omega_{k,j}+ \beta \Omega_{i,j}}. \label{eqn_final_case1_Naka2}
\end{eqnarray}
The outage probability is conditioned on the node locations (represented by the $\{\Omega_{i,j}\}$) and by the particular set of barraging transmitters (represented by $\mathcal X_j^{(t)}$).  Notice that if $|\mathcal X_j^{(t)}|= 1$, (\ref{eqn_final_case1_Naka2}) coincides with (30) in \cite{talarico:2012} and (13) in \cite{torrieri:2012}.

\section{\ac{CBR} as a Markov Process}\label{Section:MarkovChain}

Consider a single packet being transmitted through a single \ac{CBR} composed of a source (denoted $S$), a destination ($D$) and $N$ relays ($\{R_1, ...,R_N\}$). At the boundary between any two time slots, each node can be in one of the three following states, which we refer to as the {\em node state}:
\begin{itemize}
\item Node state $0$: The node has not yet successfully decoded the packet.
\item Node state $1$: It has just decoded the packet received during previous slot, and it will transmit on next slot.
\item Node state $2$: It has decoded the packet in an earlier slot and it will no longer transmit or receive that packet.
\end{itemize}
The state of the \ac{CBR} is the concatenation of the states of the individual nodes within the CBR.    Hereafter, we refer to this as the {\em \ac{CBR} state}.  The \ac{CBR} state can be compactly represented by a vector of the form $\left[S, R_1, ..., R_N, D\right]$ containing the states of the source, relays, and destination.

The behavior of the \ac{CBR} can be described as an \emph{absorbing} Markov process, which describes the dynamics of how the \ac{CBR} states evolve over a radio frame.
Define $\boldsymbol{s} = \{ s_1, s_2, \dots , s_{\varrho} \}$, which is the {\em state space} of the process composed of $\varrho$ {\em Markov} states $s_i$.   An absorbing Markov process is characterized by $\mathcal{\tau}$ {\em transient states} and $r$ {\em absorbing states}.  A state $s_i$ of a Markov chain is called \emph{absorbing} if once in that state, it is impossible to leave it, while a state that is not absorbing is called a \emph{transient} state.  We note that it is often possible to group several CBR states into a single Markov state.  For instance, we group all of the CBR states corresponding to successful decoding by the destination into a single absorbing Markov state.

\begin{figure}[t]
\centering
\includegraphics[width=9cm]{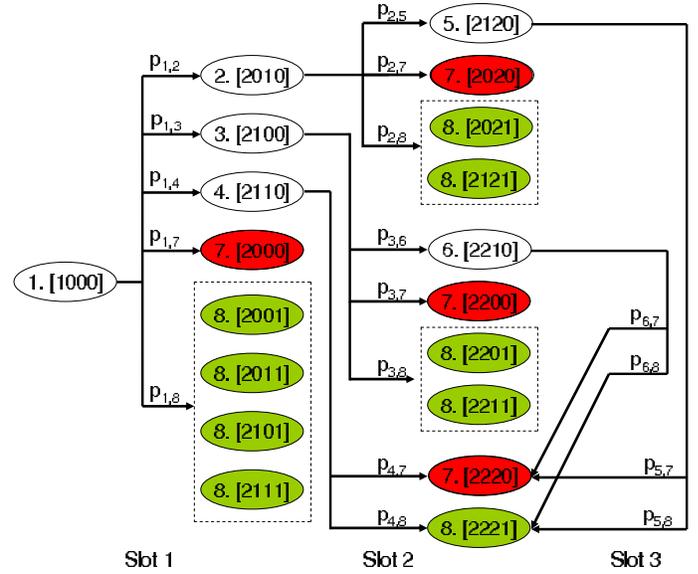}
\vspace{-0.40cm}
\caption{Markov chain for a \ac{CBR} composed of four nodes ($N=2$).  Transient states are in white, while the CBR success absorbing state is in green and the CBR outage absorbing state is in red.  Each of the two absorbing states is the union of several CBR states.}
 \label{Markov_Chain}
  \vspace{-0.60cm}
\end{figure}

The probability that the process moves from (Markov) state $s_i$ to state $s_j$ is denoted by $p_{i,j}$ and the probabilities $\{p_{i,j} \}$ are called {\em transition probabilities}.   Fig. \ref{Markov_Chain} shows the Markov chain for a CBR with $N=2$ relays (a four-node network).  The CBR states are shown as a 4-element vector, along with the transition probabilities.   The message is successfully delivered whenever the destination receives the message, which is indicated by a 1 in the last position of the CBR-state vector.  Such states are marked in green, and hereafter we refer to this condition as a {\em CBR success}.  The transmission fails whenever all entries in the CBR-state vector are either 0 or 2, indicating that none of the nodes that have not yet transmitted have successfully received the message.  Such states are marked in red, and hereafter we refer to this condition as a {\em CBR outage}.  Each transient state in the Markov process corresponds to a single CBR state, and such states are numbered 1 through 6 in the diagram.  The 4 CBR states that correspond to a CBR outage are collapsed into a single absorbing state of the Markov process (state 7), and similarly the 9 CBR states corresponding to a CBR success are collapsed into a single absorbing state of the Markov process (state 8).

Consider how the CBR state may change from time $t$ to time $t+1$.  All nodes at time $t$ with node state 1 transmit, so at time $t+1$, those nodes will now be in state 2.  All nodes that have node state 0 at time $t$ will receive the transmission, so at time $t+1$ these nodes will be either in state 1 (if the transmission was successful) or 0 (if the transmission failed).  Since the channels from transmitting to receiving nodes are independent, the state transition probability is the product of the individual transmission probabilities.  For instance, the probability of going from state [1100] to [2210] is the product of the probability that $R_2$ successfully decodes the joint transmission (from $S$ and $R_1$) and the probability that $D$ does not successfully decode the transmission.  The individual probability of successful decoding at each node is found using the outage probability of (\ref{eqn_final_case1_Naka2}).  While there is a one-to-one correspondence between CBR states and the transient Markov states, there are several CBR states associated with each of the two absorbing Markov states.  Thus, for each absorbing Markov state, the probability of transitioning into it is equal to the sum of the probabilities of transitioning into the constituent CBR states.

The Markov transition probabilities are placed into a \emph{state transition matrix} $\boldsymbol{P}$, whose $(i,j)^{th}$ entry is $\{p_{i,j}\}$. The Markov states are ordered such that the $\mathcal{\tau}$ transient states and indexed before the $r$ absorbing states.  With this ordering, the state transition matrix assumes the following canonical form \cite{Grinstead:1997}
\begin{eqnarray}
\boldsymbol{P}= \left[\begin{matrix}
  \boldsymbol{Q} & \boldsymbol{R} \\
  \boldsymbol{0} & \boldsymbol{I}
 \end{matrix}\right]
\end{eqnarray}
where $\boldsymbol{Q}$ is the $\tau \times \tau$ {\em transient matrix}, $\boldsymbol{R}$ is the $\tau \times r$ {\em absorbing matrix} and $\boldsymbol{I}$ is an $r \times r$ identity matrix.


Let $b_{i,j}$ be the probability that the process will be absorbed in the absorbing state $s_j$ if it starts in the transient state $s_i$. The {\em absorbing probability} $b_{i,j}$ is the $(i,j)^{th}$ entry of the matrix $\boldsymbol{B}$, which can be computed by (see, for instance \cite{Grinstead:1997})
\begin{eqnarray}
\boldsymbol{B}=  \boldsymbol{N} \boldsymbol{R} \label{absorbing_probabilities}
\end{eqnarray}
where $\boldsymbol{N}=\left(\boldsymbol{I}-\boldsymbol{Q} \right)^{-1}$ is the {\em fundamental matrix}.
The two absorbing states are indexed so that the first absorbing state corresponds to a CBR outage, while the second corresponds to a CBR  success. Since the process always starts in state $s_1$, it follows that $b_{1,1}$ is the {\em CBR outage probability} (which is indicated by $\epsilon_{CBR}$) and $b_{1,2}$ is the {\em CBR success probability} (which is  $\hat{\epsilon}_{CBR}= 1- \epsilon_{CBR}$).

\begin{figure}[t]
\centering
\includegraphics[width=9cm]{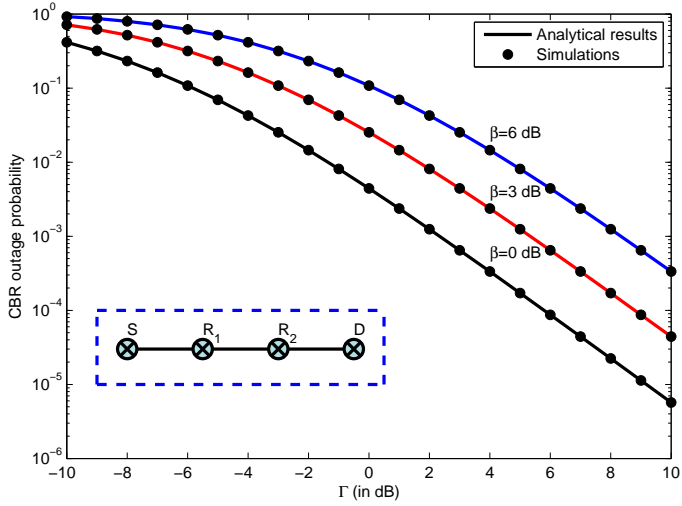}
\vspace{-0.55cm}
\caption{\ac{CBR} outage probability as a function of $\Gamma$ when \ac{CCI} is neglected. The solid lines are obtained analytically while the markers are obtained from Monte Carlo simulation.   The network topology is a four node line network shown in the inset, and $\alpha=3.5$.}
\label{EndToEnd_OutageProbability}
\vspace{-0.55cm}
\end{figure}

{\em Example $\#$ 1} Consider the four-node \ac{CBR} whose nodes are equally spaced along a line, as shown in the inset in Fig. \ref{EndToEnd_OutageProbability}.
In this example, the \ac{CCI} from adjacent \acp{CBR} is neglected, and the path-loss exponent is $\alpha=3.5$.   Fig. \ref{EndToEnd_OutageProbability} shows the CBR outage probability as a function of the SNR $\Gamma$ for three values of the threshold $\beta$. Solid curves are obtained analytically according to the methodology of this section, while the markers correspond to a simulation using the methodology introduced in \cite{Torrieri2013c}.
In particular, the simulation works by first computing the outage probabilities using (\ref{eqn_final_case1_Naka2}), using these probabilities to determine the state-transition probabilities, and simulating each state transition by drawing a random number.
The analytical results coincide with the simulations as shown in Fig. \ref{EndToEnd_OutageProbability}, and any discrepancy between the curves can be attributed to the finite number of Monte Carlo trials ($10^7$ trials were executed per \ac{SNR} point).

\section{Inter-CBR Interference}
\label{Section:IterativeMethod}



Let $\mathbf P_k$ represent the transition matrix of the $k^{th}$ CBR. The transition probabilities in $\mathbf P_k$ are computed using (\ref{eqn_final_case1_Naka2}), which depend on the $\{p_i^{(t)}\}$ associated with nodes in other CBRs.  However, each $p_i^{(t)}$ represents the probability that node $X_i$ transmits during time slot $t$, which can be found from the $\mathbf P_k$ of the corresponding CBR. Thus, interference causes a linkage between the transition probabilities in the given CBR and the transmission probabilities of adjacent CBRs.  While this linkage could be handled by considering the overall BRN as one large Markov chain,  this is a cumbersome solution as the number of states grows exponentially with the number of nodes.  A more efficient approach is an iterative one, which alternates between computing the $\{p_i^{(t)}\}$ and the $\{\mathbf P_k\}$.

Let $\mathcal S_{i,t}$ be the set of states associated with time slot $t$ and labeled with a `1' in the position corresponding to $X_i$. These are the states for which $X_i$ transmits during time slot $t$. The probability $p_i^{(t)}$ can be found by adding the probabilities of being in each of the states in $\mathcal S_{i,t}$, given that the system starts from initial state $\mathbf s_1$. The probability of being in state $\mathbf s_j  \in \mathcal S_{i,t}$ is found by multiplying the probabilities of all transitions in the Markov chain leading from state $s_1$ to state $s_j$

Let $\boldsymbol{P}_{k}[i_t]$ represent the state transition matrix for the $k^{th}$ CBR corresponding to iteration $i_t$, and similarly let  $p_i^{(t)}[i_t]$ represent the transmission probability for the $i^{th}$ user during the $t^{th}$ time slot corresponding to iteration $i_t$.    The iterative method can be described as follows:
\begin{enumerate}
\item Initialization: Set $i_t=0$ and initialize ${p^{(t)}_{i}}[0]=0, \forall i,t$, which corresponds to neglecting interference.  Then compute each $\boldsymbol{P}_{k}[0]$ using the initial $\{ p^{(t)}_{i}[0] \}$.
\item Recursion: Increment $i_t$. Evaluate each $\boldsymbol{P}_{k}[i_t]$ using the $\{ p^{(t)}_{i}[i_t-1] \}$ from the previous iteration.
\item Decision: Halt the process if $||\boldsymbol{P}^{(i_t)}_{k}-\boldsymbol{P}^{(i_t-1)}_{k}||_F<\xi, \forall k$, where the operator $||\cdot||_F$ is the Frobenius norm and $\xi$ is a tolerance.  Otherwise, go back to step 2.
\end{enumerate}

\begin{figure}[t]
\centering
\includegraphics[width=9cm]{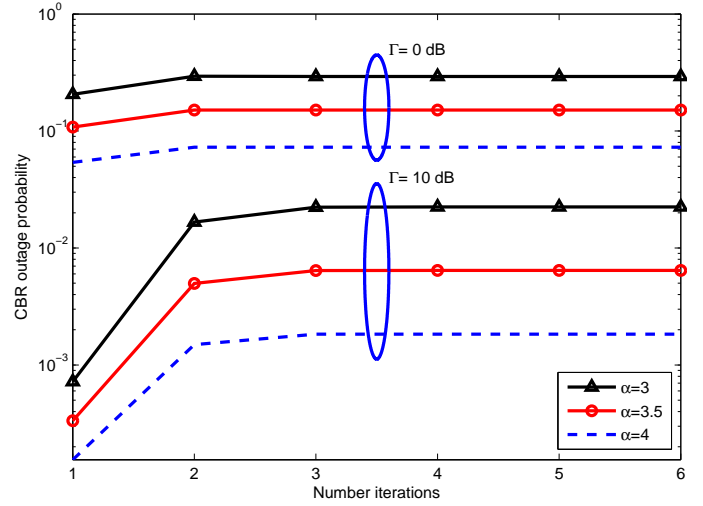}
\vspace{-0.5cm}
\caption{\ac{CBR} outage probability for the $k^{th}$ \ac{CBR} as function of the number of iterations used. Set of curves at the top: $\Gamma = 0$ dB.  Set of curves at the bottom: $\Gamma = 10$ dB. Each set of curves is obtained using three values of $\alpha$.}
 \label{Effect_Iterations}
 \vspace{-0.35cm}
\end{figure}

{\em Example $\#$ 2}   The approach is further simplified if it is assumed that the BRN is comprised of an infinite cascade of CBRs, all having identical topology.  The $\mathbf P_k$ and the set of $\{p_i^{(t)}\}$ will be the same in all active zones.  Because of this, performance can be described by a \emph{typical} CBR.
In this example, assume that each \ac{CBR} is composed of four nodes which are configured as in Example \#1. The typical \ac{CBR} is analyzed assuming it receives \ac{CCI} from just the two closest adjacent active zones (the interference from distant zones is neglected).  Fig. \ref{Effect_Iterations} shows the \ac{CBR} outage probability for a typical \ac{CBR} as a function of the number of iterations used in the algorithm described above.   Curves are shown for two values of $\Gamma$ and three values of $\alpha$.    All six curves are obtained with $\beta =6$ dB.   Fig. \ref{Effect_Iterations} shows that the \ac{CBR} outage probability increases as a function of the number of iterations, and that the iterative method converges rapidly after very few iterations.

\begin{table*}[bt]
\caption{Optimization results. }
\centering
\vspace{-0.1cm}
  \begin{tabular}{|c|c|c|c|c|c|c|c|}
  \hline
   $\Gamma$ (in dB) & \ac{CCI} & $R$ & $N$ & $d$ & Optimal nodes' location for a line network & $\Upsilon_{\mathsf{opt}}$\\
  \hline
  \hline
  0        &   \xmark          &  4.452    &   0   &  0.3 & \includegraphics[scale=0.535]{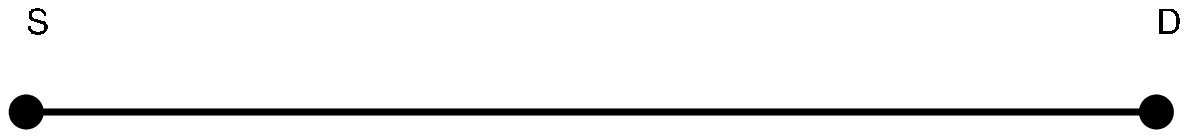}  & 0.490  \\
  \cline{3-7}
    &           \cmark           &  4.421    &   5   & 1.2 & \includegraphics[scale=0.535]{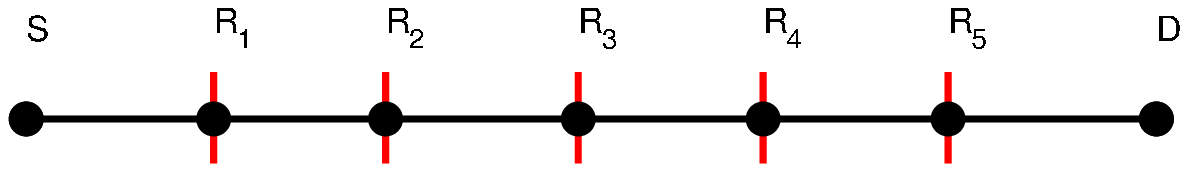}& 0.402 \\
  \hline
  \hline
  5     &   \xmark          &   4.611   &   0   &  0.4 & \includegraphics[scale=0.535]{Topology1}  & 0.683  \\
  \cline{3-7}
           & \cmark           &   4.452    &   5   & 1.7 & \includegraphics[scale=0.535]{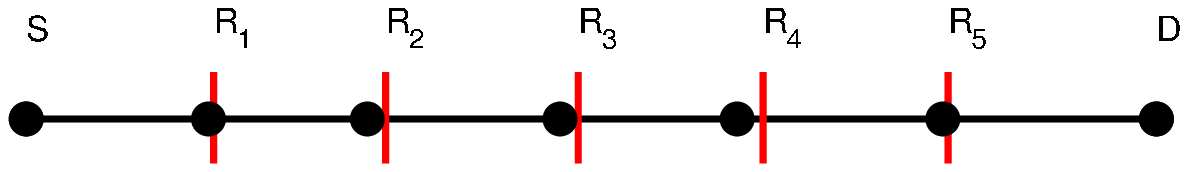}& 0.559 \\
  \hline
  \hline
   10       &   \xmark          &  5.028    &   0   &  0.5 & \includegraphics[scale=0.535]{Topology1}  & 0.951  \\
  \cline{3-7}
            & \cmark           &  4.547   &   5   & 2.3 & \includegraphics[scale=0.535]{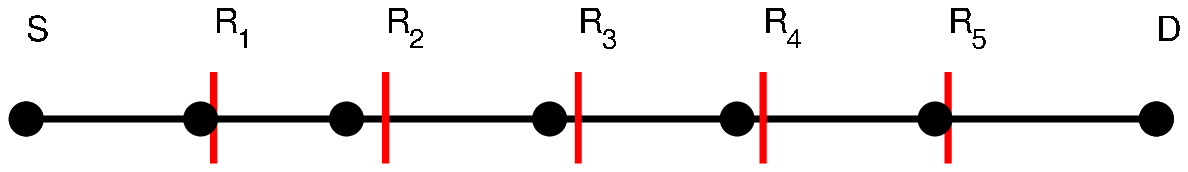}& 0.777 \\
  \hline
  \end{tabular}
    \label{main_table}
    \vspace{-0.35cm}
\end{table*}

\section{Network Optimization}
\label{Section:Optimization}


The \acl{TC} \cite{Gupta2000} of an ad hoc network quantifies the rate that data can be reliably communicated over a unit of distance, and is typically expressed in units of meter-bits-per-second. In the context of a \ac{BRN}, the \ac{TC} of a typical  \ac{CBR} is
\begin{eqnarray}
 \Upsilon=  T d
\label{trasport capacity}
\end{eqnarray}
where $T$ is the \emph{throughput} of the \ac{CBR} and $d$ is the distance between the \ac{CBR}'s source and destination. From \cite{ghanim2006}, the throughput can be written as
\vspace{0.1 cm}
\begin{eqnarray}
T=\frac{(1-\epsilon_{CBR})}{2 F} R
\label{throughput}
\end{eqnarray}
where $R$ is the code rate expressed in units of bits per channel use (bpcu) and the factor 2 in the denominator is a consequence of the buffer nodes operating in half-duplex mode, or, equivalently, due to alternating between active and silent zones.

Substituting (\ref{throughput}) into (\ref{trasport capacity}), and relating the rate and the SINR threshold by $R=\log_2\left( 1+\beta \right)$, which is the Shannon capacity for complex discrete-time AWGN channels, yields
\begin{eqnarray}
\Upsilon=  d\frac{\hat{\epsilon}_{CBR}}{2 \left(N+1\right)}\log_2\left( 1+\beta \right).
\label{tau}
\end{eqnarray}

Let $\boldsymbol{X}_k$ be a vector containing the position of the $N$ relays in the $k^{th}$ CBR. The goal of the network optimization is to determine the set $\boldsymbol{\Theta}=(\boldsymbol{X}_k, R, N, d)$ that maximizes the \ac{TC}.
Because the \ac{TC} is nonlinear in $\boldsymbol{\Theta}$ and the search space is multi-dimensional, efficiently finding a solution is a challenge.  However, by initially using an exhaustive search, we have confirmed that the optimization surface is convex over the set $(R, N, d)$.  Thus, the optimization can be solved through a convex optimization over $(R, N, d)$ combined by a stochastic optimization over $\boldsymbol{X}_k$.  It follows that an efficient approach to finding the optimal set $\boldsymbol{\Theta}$ is as follows:

\begin{enumerate}
 \item For each of the parameters $N$, $R$ and $d$, select a pair of endpoints and a midpoint ($3^3 = 27$ points).  The endpoints should initially be far enough apart to guarantee that the optimal point lies within the endpoints.
 \item Define a {\em reference vector} $\boldsymbol{X}_\mathsf{ref}\left(N,d\right)$, which is a vector $\boldsymbol{X}_k$ for the pair $(N,d)$ and a {\em reference set} $\{\boldsymbol{X}_\mathsf{ref}\left(N,d\right)\}$ containing the reference vectors of all $(N,d)$ considered by the algorithm.
     Create three vectors $\{\boldsymbol{X}_k\}$ for each pair of $(N,d)$.
     When $(N,d)$ is chosen for the first time, the first vector of $\{\boldsymbol{X}_k\}$ is obtained by placing the $N$ relays on the length-$d$ line connecting the source and the destination and this vector is added to the reference set, constituting $\boldsymbol{X}_\mathsf{ref}\left(N,d\right)$ for the pair $(N,d)$ chosen. The placement is done such that relays are separated by a minimum distance $r_\mathsf{s}=\frac{d}{3N}$, but are otherwise uniformly distributed along the line.
     If the pair $(N,d)$ has been already used, the first vector of $\{\boldsymbol{X}_k\}$ is set to be equal to $\boldsymbol{X}_\mathsf{ref}\left(N,d\right)$.
     The other two vectors are obtained by {\em mutating} $\boldsymbol{X}_\mathsf{ref}\left(N,d\right)$ as follows:
     \begin{enumerate}
     \item For each relay in $\boldsymbol{X}_\mathsf{ref}\left(N,d\right)$, draw a Bernoulli random variable indicating if the relay should be moved or if it should remain in its current location.
     \item If the node $X_i$ is to be moved, the direction (right or left) of movement is selected with equal probability.  The new position is obtained by moving the current position in the selected direction by a distance $\Delta=\frac{d}{n_{\Delta} N}$, where initially $n_{\Delta}=2$.
     \end{enumerate}
 \label{step1}
 \item Compute the TC for the endpoints and the midpoints of the interval of one of the parameters in the set $(R, N, d)$ and for the different sets of locations of the relays. For each pair $(N,d)$ chosen, determine the vector $\boldsymbol{X}_k$ that provides the higher TC and replace the $\boldsymbol{X}_\mathsf{ref}\left(N,d\right)$ in the reference set with this vector.
     \label{step2}
 \item For the same parameter of the set $(R, N, d)$ used in step \ref{step2}, move the midpoint of the interval towards the endpoint that gives higher TC. For each pair $(N,d)$, generate the three sets of positions as in step \ref{step1}. \label{step3}
 \item Repeat step \ref{step2} and \ref{step3} recursively for all the parameters in the set $(R, N, d)$ and gradually reduce for one parameter at the time the range between the endpoints and the midpoint until a certain tolerance is reached. If the optimal TC remains the same as in the previous iteration $n_{\Delta}$ is increased by one, until a certain value is achieved.
 \item If the optimal TC remains the same as in the previous iteration, for all the parameters in the set $(R, N, d)$ it is not possible to further reduce the range between the endpoints and the midpoint, since a given tolerance is achieved, and $n_{\Delta}$ reaches its maximum value, the algorithm stops.
 \end{enumerate}

Using the methodology described above, optimization results were obtained under the same assumptions that were used in Example $\#$2. In particular, it is assumed that the \ac{BRN} extends infinitely and \acp{CBR} are composed of line networks with the same number of relays which are all equally positioned. For each \ac{CBR} the only \ac{CCI} considered are the ones from the two closest adjacent active zones (the interference from farther active zones is neglected, since they are at least at $3d$ distance away). Although this assumptions can be relaxed using the analysis and methodology proposed along this paper, in this section they are used for simplicity of exposure.

Table \ref{main_table} shows the optimal values of $N$, $R$, $d$ and the optimal location for the $N$ relays. Optimization results are provided for three values of $\Gamma$ and when the \ac{CCI} is neglected as well as when they are taken into account.
Vertical red lines indicate the optimal position of the mobiles for $\Gamma=0$ dB, and they are used to facilitate the graphical analysis and comparison of the optimal location of the mobiles for different scenarios. For all scenarios considered, the path loss exponent is fixed to $\alpha=3.5$. Table \ref{main_table} emphasizes that when there is no \ac{CCI}, the optimal configuration for a cooperative \ac{BRN} is to use short \acp{CBR} characterized by a point-to-point conventional communication, while five relays distributed over a larger \ac{CBR} allow to achieve the optimal \ac{TC} when there is \ac{CCI} from the adjacent \acp{CBR}. As expected, the optimal \ac{TC} increases by increasing $\Gamma$, since the links experience a more favorable channel. Furthermore, the optimal \ac{TC} increases going from a scenario in which there is \ac{CCI} to one in which \ac{CCI} is neglected, and this is more prominent as $\Gamma$ increases since the network becomes more interference sensitive. In agreement with \cite{chattopadhyay:2012}, an increase in $\Gamma$ produces a shift towards the source of the optimal position of the relays, which increases the likelihood to cooperate among themselves. A more favorable condition of the channel produces an increment in both the optimal code rate $R$ and in the optimal dimension of the \ac{CBR}.

\balance

\section{Conclusion} \label{Section:Conclusion}

This paper presents a new analysis and optimization for unicast in a \ac{BRN}. A \ac{BRN} is analyzed by describing the behavior of each constituent \acp{CBR} as a Markov process. The transition probabilities are computed by using a new closed form expression for the outage probability, which takes into account the path loss, Rayleigh fading, and interference. Interference causes an interdependence between the transmission and transition probabilities, which is taken into account by an iterative method, that updates the probability of collisions for each \ac{CBR} at each iteration for each one of the time slots. The analysis is used to optimize a \ac{BRN} and in particular to maximize the \ac{TC} by finding the optimal number of relays $N$ that need to be used in the \acp{CBR} that compose the \ac{BRN}, their optimal placement $\boldsymbol{X}_k$, the optimal size of the \acp{CBR} $d$ and the optimal code rate $R$ at which the nodes transmit. While the analysis and the model have been used to study and optimize a \ac{BRN}, this  work  could  be  extended to different types of cooperative ad hoc networks, for which multiple source transmissions are diversity combined at the receiver.

\balance

\bibliographystyle{ieeetr}
\bibliography{Barrage2014}

\end{document}